\documentclass[showpacs,fleqn,twocolumn,nobibnotes]{revtex4}

\usepackage{amsmath}
\usepackage{graphicx}

\def\lsim{\raise0.3ex\hbox{$<$\kern-0.75em\raise-1.1ex\hbox{$\sim$}}}
\def\gsim{\raise0.3ex\hbox{$>$\kern-0.75em\raise-1.1ex\hbox{$\sim$}}}

\def\pom{{I\!\!P}}

\newcommand{\rr}{\mbox{\boldmath $r$}}

\newcommand{\rb}{\mbox{\boldmath $b$}}
\newcommand{\rd}{\mbox{\boldmath $\Delta$}}

\begin{document}

\title{Non-linear QCD dynamics and exclusive production in $ep$ collisions}
\author{V.P. Gon\c{c}alves
$^{a}$, M.V.T. Machado $^{b}$ and A.R. Meneses$^{a}$}

\affiliation{$^a$ Instituto de F\'{\i}sica e Matem\'atica, Universidade Federal de
Pelotas\\
Caixa Postal 354, CEP 96010-900, Pelotas, RS, Brazil.\\
$^b$ Centro de Ci\^encias Exatas e Tecnol\'ogicas, Universidade Federal do Pampa \\ Campus de Bag\'e, Rua Carlos Barbosa,  CEP 96400-700, Bag\'e, RS,  Brazil.}

\begin{abstract}
The exclusive processes in electron-proton ($ep$) interactions  are an important tool to investigate the QCD dynamics at high energies as they are in general driven by the gluon content of proton which is strongly subject to parton saturation effects.  In this paper we compute the cross sections for the exclusive vector meson production as well as the deeply virtual Compton scattering (DVCS) relying on the color dipole approach and considering the numerical solution of the  Balitsky-Kovchegov equation including running coupling corrections. We show that the small-$x$ evolution given by this evolution equation  is able to describe the DESY-HERA data and is relevant for the physics of the exclusive observables in  future electron-proton colliders  and in photoproduction processes to be measured in coherent interactions at the LHC.

\end{abstract}

\maketitle
\section{Introduction}
The understanding of the high energy (small $x$) regime of Quantum Chromodynamics (QCD) has been one of the main challenges of this theory,
which has been intensely investigated through high energy collision experiments.
This regime, where one expects to observe the non-linear behavior predicted
by theoretical developments, has been explored in $ep$ collisions at DESY-HERA and $pp/dA$ collisions at BNL-RHIC and, in a near future, in $pp/pA/AA$ collisions at CERN-LHC. In particular, at high energies, the growth of the parton distribution is expected to saturate, forming a  Color Glass Condensate (CGC), whose evolution with energy is described by an infinite hierarchy of coupled equations for the correlators of  Wilson lines \cite{BAL,CGC}.  
In the mean field approximation, the first equation of this  hierarchy decouples and boils down to a single non-linear integro-differential  equation: the Balitsky-Kovchegov (BK) equation \cite{BAL,KOVCHEGOV}. This equation determines, in the large-$N_c$ (the number of colors) limit, the evolution of the two-point correlation function, which corresponds to the  scattering amplitude ${\cal{N}}(x,r,b)$ of a dipole off the CGC, where $r$ is the dipole size and $b$ the impact parameter. This quantity  encodes the information about the hadronic scattering  and then about the non-linear and quantum effects in the hadron wave function (For recent reviews, see e.g. \cite{hdqcd}). 
Recently, the next-to-leading order corrections to BK equation were calculated  \cite{kovwei1,javier_kov,balnlo} through the ressumation of $\alpha_s N_f$ contributions to all orders, where $N_f$ is the number of flavors. Such calculation allows one to estimate the soft gluon emission and running coupling corrections to the evolution kernel and, in particular, the authors have verified that  the dominant contributions come from the running coupling corrections, which allows to determine the scale of the running coupling in the kernel. The solution of the improved BK equation was studied in detail in Refs. \cite{javier_kov,javier_prl}. Basically, one has that the running of the coupling reduces the speed of the evolution to values compatible with experimental data, with the geometric scaling regime being reached only at ultra-high energies. In Ref. \cite{javier_prl}, the solution of the improved BK equation was used to calculate the pseudorapidity density of charged particles produced in nucleus-nucleus collisions and a remarkable good agreement with the RHIC data was observed. More recently, a global analysis of the small $x$ data for the proton structure function using the improved BK equation was performed \cite{bkrunning} (See also Ref. \cite{weigert}). In contrast to the  BK  equation at leading logarithmic $\alpha_s \ln (1/x)$ approximation, which  fails to describe data, the inclusion
of running coupling effects to evolution renders BK equation compatible with them. The improved BK equation has been shown to
be really successful when applied to the description of the $ep$ HERA data for the inclusive and diffractive proton structure function \cite{bkrunning,weigert,vic_joao}, as well as for the  forward hadron  spectra in $pp$ and $dA$ collisions \cite{vic_joao,alba_marquet},  which motivates us to extend the study for exclusive observables.


Exclusive processes in deep inelastic scattering (DIS) have appeared as key reactions to trigger the generic mechanism of diffractive scattering. 
In particular, the diffractive vector meson production and deeply virtual Compton scattering (DVCS) have been extensively studied at HERA and provide a valuable  probe of the  QCD dynamics at high energies. In a general way, these processes are driven by the gluon content of target (proton or nuclei) which is strongly subject to parton saturation effects as well as considerable nuclear shadowing corrections when one considers scattering on nuclei. In particular, the cross section for exclusive processes in DIS are proportional to the square of scattering amplitude, which turn it  strongly sensitive to the underlying QCD dynamics. They have been successfully described using color dipole approach and phenomenological model inspired in general aspects of parton saturation physics  \cite{teaney,KMW,MPS}.  Here, we will make use of numerical solution of the  Balitsky-Kovchegov equation including running coupling corrections in order to estimate the contribution of the saturation physics for exclusive processes. Our analysis is  relevant for the physics to be studied  in future electron - proton collider, as e.g. the LHeC \cite{dainton},  and in photoproduction processes in coherent interactions at the LHC \cite{nos}. This paper is organized as follows. In next section (Section \ref{exc}) the main formula for computing the differential cross section for exclusive processes in DIS are presented. Moreover, we discuss the main aspects of the running corrections for the BK equation. In Section \ref{results} we present our results and discussions.

\section{Exclusive processes in DIS and the RC BK solution}
\label{exc}

Let us consider photon-hadron scattering in the dipole frame, in which most of the energy is
carried by the hadron, while the  photon  has
just enough energy to dissociate into a quark-antiquark pair
before the scattering. In this representation the probing
projectile fluctuates into a
quark-antiquark pair (a dipole) with transverse separation
$\rr$ long after the interaction, which then
scatters off the hadron \cite{nik}.
In the dipole picture the   amplitude for production of an exclusive final state $E$, such as a vector meson ($E = V$) or a real photon in DVCS ($E = \gamma$) is given by (See e.g. Refs. \cite{nik,vicmag_mesons,KMW})
\begin{eqnarray}
\, {\cal A}_{T,L}^{\gamma^*p \rightarrow E p}\, (x,Q^2,\Delta)  = 
\int dz\, d^2\rr \,(\Psi^{E*}\Psi)_{T,L}\,{\cal{A}}_{q\bar{q}}(x,\rr,\Delta) \, ,
\label{sigmatot}
\end{eqnarray}
where $(\Psi^{E*}\Psi)_{T,L}$ denotes the overlap of the photon and exclusive final state wave functions. The variable  $z$ $(1-z)$ is the
longitudinal momentum fractions of the quark (antiquark),  $\Delta$ denotes the transverse momentum lost by the outgoing proton ($t = - \Delta^2$) and $x$ is the Bjorken variable. For DVCS, the amplitude involves a sum over quark flavors. Moreover, ${\cal{A}}_{q\bar{q}}$ is the elementary elastic amplitude for the scattering of a dipole of size $\rr$ on the target. It is directly related to ${\cal{N}} (x,\rr,\rb)$ and consequently to the QCD dynamics (see below). One has that \cite{KMW}
\begin{eqnarray}
{\cal{A}}_{q\bar{q}} (x,\rr,\Delta) & = & i \int d^2 \rb \, e^{-i \rb.\rd}\, 2 {\cal{N}}(x,\rr,\rb) \,\,,
\end{eqnarray}
where $\rb$ is the transverse distance from the center of the target to one of the $q \bar{q}$ pair of the dipole.  Consequently, one can express the amplitude for the exclusive production of a final state $E$ as follows
\begin{eqnarray}
 {\cal A}_{T,L}^{\gamma^*p \rightarrow E p}(x,Q^2,\Delta) & = & i
\int dz \, d^2\rr \, d^2\rb  e^{-i[\rb-(1-z)\rr].\rd} \nonumber \\
 &\times & (\Psi_{E}^* \Psi)_T \,2 {\cal{N}}(x,\rr,\rb)
\label{sigmatot2}
\end{eqnarray}
where  the factor $[i(1-z)\rr].\rd$ in the exponential  arises when one takes into account non-forward corrections to the wave functions \cite{non}.
Finally, the differential cross section  for  exclusive production is given by
\begin{eqnarray}
\frac{d\sigma_{T,L}}{dt} (\gamma^* p \rightarrow E p) = \frac{1}{16\pi} |{\cal{A}}_{T,L}^{\gamma^*p \rightarrow E p}(x,Q^2,\Delta)|^2\,(1 + \beta^2)\,,
\label{totalcs}
\end{eqnarray}
where $\beta$ is the ratio of real to imaginary parts of the scattering
amplitude. For the case of heavy mesons, skewness corrections are quite important and they are also taken  into account. (For details, see Refs. \cite{vicmag_mesons,KMW} and Section \ref{results}).


The photon wavefunctions appearing in Eq. (\ref{sigmatot2}) are well known in literature \cite{KMW}. For the meson wavefunction, we have considered the Gauss-LC  model \cite{KMW} which is a simplification of the DGKP wavefunctions. The motivation for this choice is its simplicity and the fact that the results are not sensitive to a different model. In photoproduction, this leads only to an  uncertainty  of a few percents in overall normalization. We consider the quark masses $m_{u,d,s} = 0.14$ GeV, $m_c = 1.4$ GeV and $m_b=4.5$ GeV. The parameters for the meson wavefunction can be found in Ref. \cite{KMW}. In the DVCS case, as one has a real photon at the initial state, only the transversely polarized overlap function contributes to the cross section.  Summed over the quark helicities, for a given quark flavor $f$ it is given by \cite{MW},
\begin{eqnarray}
  (\Psi_{\gamma}^*\Psi)_{T}^f & = & \frac{N_c\,\alpha_{\mathrm{em}}e_f^2}{2\pi^2}\left\{\left[z^2+\bar{z}^2\right]\varepsilon_1 K_1(\varepsilon_1 r) \varepsilon_2 K_1(\varepsilon_2 r) \right. \nonumber \\
& + & \left.   m_f^2 K_0(\varepsilon_1 r) K_0(\varepsilon_2 r)\right\},
  \label{eq:overlap_dvcs}
\end{eqnarray}
where we have defined the quantities $\varepsilon_{1,2}^2 = z\bar{z}\,Q_{1,2}^2+m_f^2$ and $\bar{z}=(1-z)$. Accordingly, the photon virtualities are $Q_1^2=Q^2$ (incoming virtual photon) and $Q_2^2=0$ (outgoing real photon).

The scattering amplitude ${\cal{N}}(x,\rr,\rb)$   contains all
information about the target and the strong interaction physics.
In the Color Glass Condensate (CGC)  formalism \cite{CGC,BAL}, it  encodes all the
information about the
non-linear and quantum effects in the hadron wave function. It can be obtained by solving an appropriate evolution
equation in the rapidity $y\equiv \ln (1/x)$, which in its  simplest form is the Balitsky-Kovchegov equation.  In leading
order (LO), and in the translational invariance approximation---in which the scattering
amplitude does not depend on the collision impact parameter $\bm{b}$---it reads

	\begin{eqnarray}\label{eq:bklo}
		\frac{\partial {\cal{N}}(r,Y)}{\partial Y} & = & \int {\rm d}\bm{r_1}\, K^{\rm{LO}}
		(\bm{r,r_1,r_2})
		[{\cal{N}}(r_1,Y)+{\cal{N}}(r_2,Y)\nonumber \\
& - & {\cal{N}}(r,Y)-{\cal{N}}(r_1,Y){\cal{N}}(r_2,Y)],
	\end{eqnarray}
where ${\cal{N}}(r,Y)$ is the scattering amplitude for a dipole (a quark-antiquark pair)
off a target, with transverse size $r\equiv |\bm{r}|$, $Y\equiv \ln(x_0/x)$ ($x_0$ is the value of $x$ where the evolution starts), and $\bm{r_2 = r-r_1}$. $K^{\rm{LO}}$ is the evolution kernel, given by

	\begin{equation}\label{eq:klo}
		K^{\rm{LO}}(\bm{r,r_1,r_2}) = \frac{N_c\alpha_s}{2\pi^2}\frac{r^2}{r_1^2r_2^2},
	\end{equation}
where $\alpha_s$ is the (fixed) strong coupling constant. This equation is a
generalization of the linear BFKL equation (which corresponds of the first three terms), with the inclusion
of the (non-linear) quadratic term, which damps the indefinite growth of the amplitude
with energy predicted by BFKL evolution. It has been shown \cite{mp} to be in the same
universality class of the Fisher-Kolmogorov-Pertovsky-Piscounov (FKPP) equation
\cite{fkpp} and, as a consequence, it admits the so-called traveling wave solutions.
This means that, at asymptotic rapidities, the scattering amplitude is a wavefront which
travels to larger values of $r$ as $Y$ increases, keeping its shape unchanged. Thus,
in such asymptotic regime, instead of depending separately on $r$ and $Y$, the amplitude
depends on the combined variable $rQ_s(Y)$, where $Q_s(Y)$ is the saturation scale. This property of the solution of BK equation is a natural
explanation to the {\it geometric scaling}, a phenomenological feature observed at the
DESY $ep$ collider HERA, in the measurements of inclusive and exclusive processes
\cite{scaling,marquet,prl,prl1}. Although having its properties been intensely studied and understood, both numerically
and analytically, the LO BK equation presents some difficulties when applied to study
DIS small-$x$ data. In particular, some studies concerning this equation
\cite{IANCUGEO,MT02,AB01,BRAUN03,AAMS05} have
shown that the resulting saturation scale grows much faster with increasing energy
($Q_s^2\sim x^{-\lambda}$, with $\lambda\simeq 4.88N_c\alpha_s/\pi \approx 0.5$ for  $\alpha_s = 0.2$) than that
extracted from phenomenology ($\lambda \sim 0.2-0.3$). This difficulty could be solved by
considering smaller values of the strong coupling constant $\alpha_s$, but this procedure
would lead to physically unrealistic values. One can conclude that higher order corrections
to LO BK equation should be taken into account to make it able to describe the available
small-$x$ data.

\begin{figure}[t]
\includegraphics[scale=0.35]{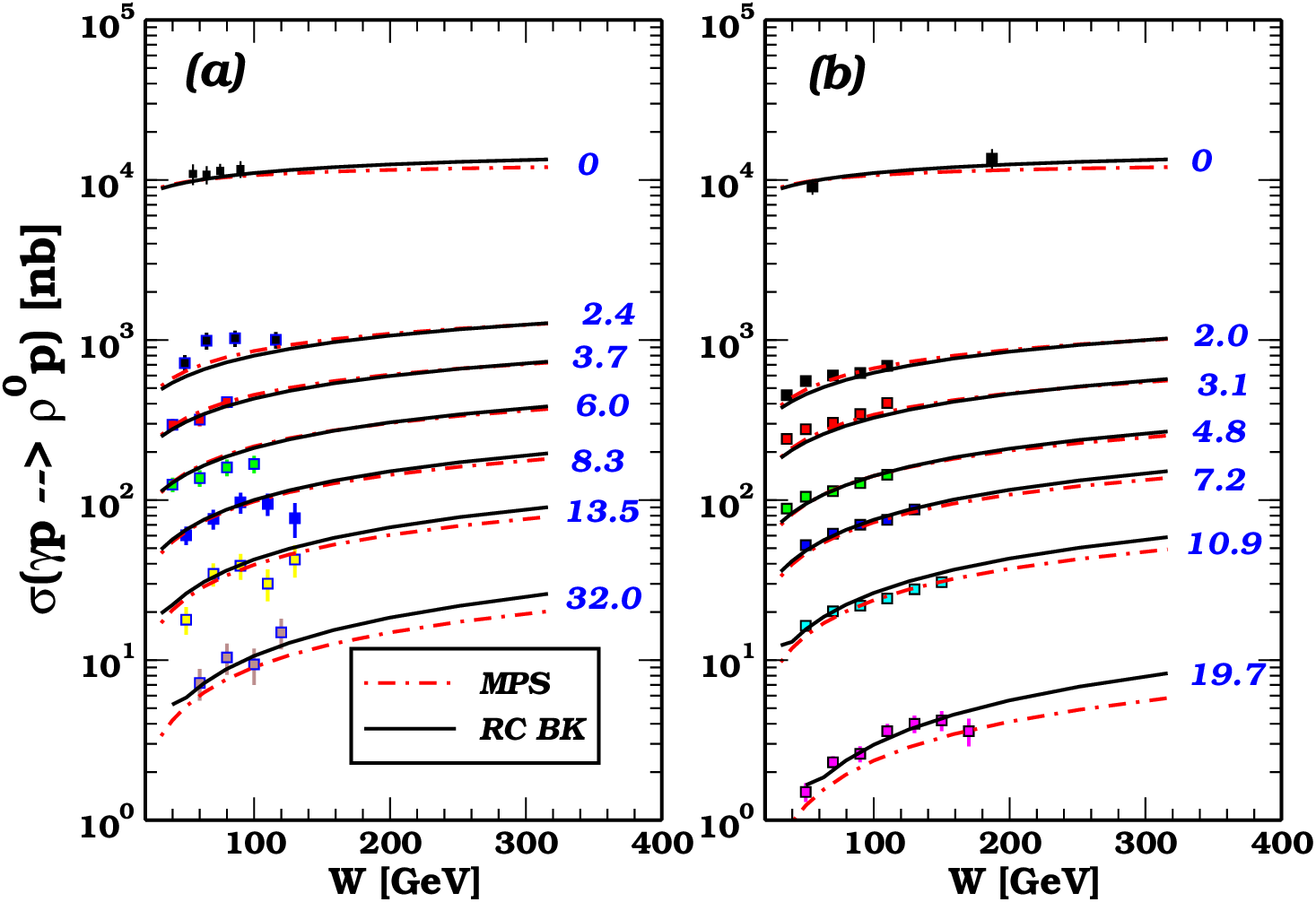}
\caption{(Color online)  Energy dependence of the $\gamma p$ cross section for $\rho^0$ production  for different photon virtualities. Data from (a) ZEUS and (b) H1 collaborations \cite{H1_rho,ZEUS_rho}.}
\label{fig:1}
\end{figure}

\begin{figure}[t]
\includegraphics[scale=0.35]{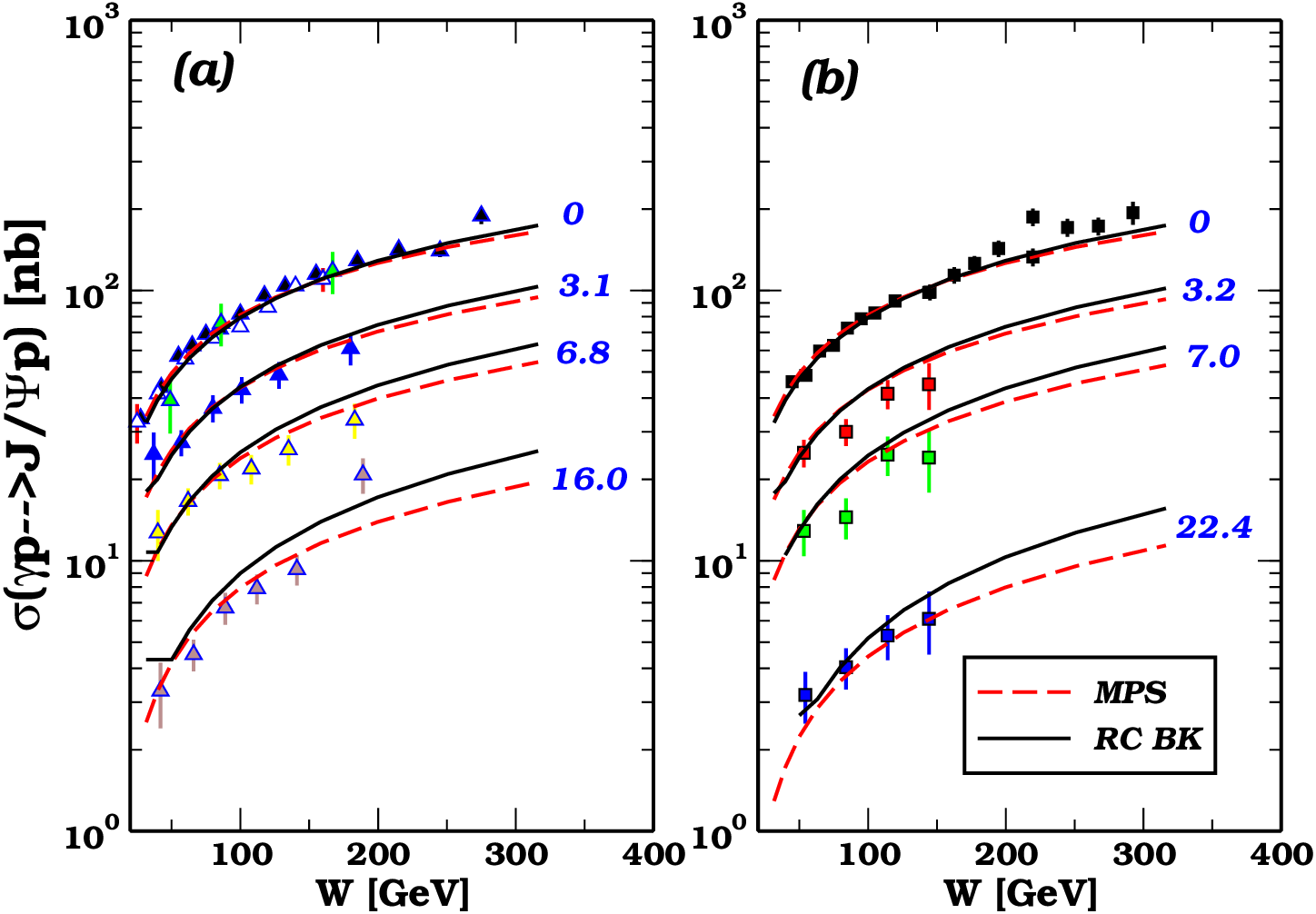}
\caption{(Color online)  Energy dependence of the $\gamma p$ cross section for $J/\Psi$ production  for different photon virtualities. Data from (a) ZEUS and (b) H1 collaborations \cite{ZEUS_jpsi, H1_jpsi}.}
\label{fig:2}
\end{figure}

The calculation of the running coupling corrections to BK evolution kernel was explicitly
performed in \cite{kovwei1,balnlo}, where the authors included $\alpha_sN_f$ corrections to the kernel to all orders. The  improved  BK equation is given in terms of a  
running coupling and a subtraction term, with the latter accounting for conformal, non running coupling contributions. In the prescription proposed by Balitsky in \cite{balnlo} to single out the ultra-violet divergent contributions from the finite ones that originate after the resummation of quark loops, the contribution of the subtraction term is minimized at large energies. In \cite{bkrunning} this contribution was disregarded, and the improved BK equation was numerically solved replacing the leading order kernel  in Eq. (\ref{eq:bklo}) by the modified kernel which includes the running coupling
corrections and  is given by \cite{balnlo}
	\begin{eqnarray}\label{eq:krun}
		K^{\rm{Bal}}(\bm{r,r_1,r_2}) & = & \frac{N_c\alpha_s(r^2)}{2\pi^2}
		\left[\frac{r^2}{r_1^2r_2^2}  
              +  \frac{1}{r_1^2}\left(\frac{\alpha_s(r_1^2)}
		{\alpha_s(r_2^2)}-1\right) \right. \nonumber \\
& + & \left.  \frac{1}{r_2^2}\left(\frac{\alpha_s(r_2^2)}
		{\alpha_s(r_1^2)}-1\right)\right] .
	\end{eqnarray}
From a recent numerical study of the improved BK equation \cite{javier_kov}, it has been confirmed that the running coupling corrections lead to a considerable
increase in the anomalous dimension and to a slow-down of the evolution
speed, which implies, for example, a slower growth of the saturation scale with
energy, in contrast with the faster growth predicted by the LO BK equation. Moreover, 
as shown in \cite{bkrunning,vic_joao,alba_marquet} the improved BK equation has been shown to
be really successful when applied to the description of the $ep$ HERA data for the inclusive and diffractive proton structure function, as well as for the  forward hadron  spectra in $pp$ and $dA$ collisions. 
It is important to emphasize that the impact parameter dependence was not taken into account in Ref. \cite{bkrunning}, the normalization of the dipole cross section was fitted to data and two distinct initial conditions, inspired in the Golec Biernat-Wusthoff (GBW) \cite{GBW} and McLerran-Venugopalan (MV) \cite{MV} models, were considered. The predictions resulted to be almost independent of the initial conditions and, besides, it was observed that it is impossible to describe the experimental data using only the linear limit of the BK equation, which is equivalent to Balitsky-Fadin-Kuraev-Lipatov (BFKL) equation \cite{bfkl}. In next section we will compare the results of the RC BK approach to the experimental data on exclusive processes at DESY-HERA and present our predictions for the kinematical range of the future electron - proton collider \cite{dainton}.

\section{Results and discussions}
\label{results}

In what follows we calculate the exclusive observables using as input in our calculations the solution of the RC BK evolution equation. In particular, we make
use of the public-use code available in \cite{code}. In numerical calculations we have considered the GBW initial condition for the evolution (we quote Ref. \cite{bkrunning}  for details) and it was verified the MV initial condition gives cross section with overall normalization $10-15\,\%$ smaller and unchanged energy dependence. Furthermore, we compare the RC BK predictions with those from the
non-forward saturation model of Ref. \cite{MPS} (hereafter MPS model), which captures the main features of the dependence on energy,  virtual photon virtuality and momentum transfer $t$.  In the MPS model, the elementary elastic amplitude for dipole interaction is given by,
\begin{eqnarray}
\label{sigdipt}
\mathcal{A}_{q\bar q}(x,r,\Delta)= 2\pi R_p^2\,e^{-B|t|} {\cal{N}} \left(rQ_{\mathrm{sat}}(x,|t|),x\right),
\end{eqnarray}
with the asymptotic behaviors $Q_{\mathrm{sat}}^2(x,\Delta)\sim
\max(Q_0^2,\Delta^2)\,\exp[-\lambda \ln(x)]$. Specifically, the $t$ dependence of the saturation scale is parametrised as
\begin{eqnarray}
\label{qsatt}
Q_{\mathrm{sat}}^2\,(x,|t|)=Q_0^2(1+c|t|)\:\left(\frac{1}{x}\right)^{\lambda}\,, \end{eqnarray}
in order to interpolate smoothly between the small and intermediate transfer
regions. For the parameter $B$ we use the value $B=3.754$ GeV$^{-2}$ \cite{MPS}. Finally, the scaling function ${\cal{N}}$ is obtained from the forward saturation model \cite{IIM}. 

Here, in order to take into account the skewedness correction, in the limit that $x^\prime \ll x \ll 1$, the elastic differential cross section should be multiplied by a factor $R_g^2$, given by \cite{Shuvaev:1999ce}
\begin{eqnarray}
\label{eq:Rg}
  R_g(\lambda_e) = \frac{2^{2\lambda_e+3}}{\sqrt{\pi}}\frac{\Gamma(\lambda_e+5/2)}{\Gamma(\lambda_e+4)}, \nonumber \\
  \quad\text{with} \quad \lambda_e \equiv \frac{\partial\ln\left[\mathcal{A}(x,\,Q^2,\,\Delta)\right]}{\partial\ln(1/x)},
\end{eqnarray}
which gives an important contribution mostly at large virtualities. In addition, we will take into account the correction for real part of the amplitude, using dispersion relations $Re {\cal A}/Im {\cal A}=\mathrm{tan}\,(\pi \lambda_e/2)$. In the MPS model, the skewedness correction is absorbed in the model parameters and only real part of amplitude will be considered.

\begin{figure}[t]
\includegraphics[scale=0.35]{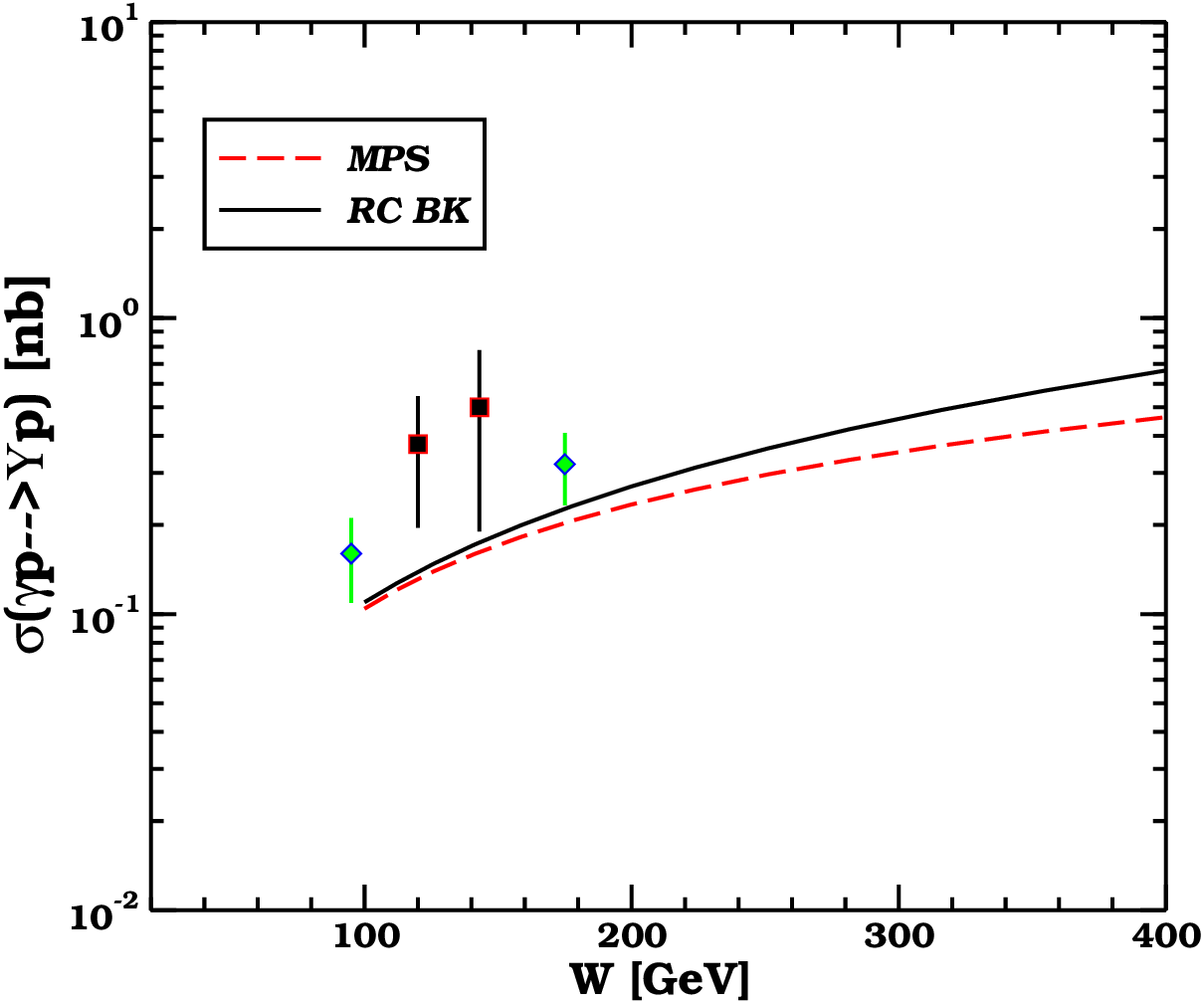}
\caption{(Color online)  Energy dependence of the $\gamma p$ cross section for $\Upsilon$ photoproduction. Data from ZEUS and H1 collaborations \cite{ZEUS_ups,H1_ups}.}
\label{fig:3}
\end{figure}


Let us start to compare the RC BK predictions to the available HERA data for exclusive vector meson ($\rho$, $J/\Psi$ and $\Upsilon$) photo and electroproduction. In Fig. \ref{fig:1} we present the predictions of the RC BK and MPS models  for the diffractive $\rho^0$ vector meson production  and compare it with the current experimental data from ZEUS (left panel)  and H1 (right panel) Collaborations \cite{H1_rho,ZEUS_rho}.  These measurements are  interesting as they cover momenta scale  that are in the transition region between perturbative and nonperturbative physics, where saturation effects is expected to play an very important role. As the numerical RC BK solution there exists only for forward dipole-target amplitude we need an approximation to compute the non-forward amplitude. Here, we assume the usual exponential ansatz for the $t$-dependence which implies that the total cross-section is given by
\begin{eqnarray}
\sigma_{tot}(\gamma^*p\rightarrow Vp) = \frac{1}{B_V}\,\left. \left. \left[\frac{d\sigma_T}{dt}\right|_{t=0} + \frac{d\sigma_L}{dt}\right|_{t=0}\right]\,.
\end{eqnarray}
Notice that values of the slope parameter $B_V$ are not very accurately measured. We use the parametrisation
\begin{eqnarray}
B_V\,(Q^2) = 0.60\,\left[ \frac{14}{(Q^2+M_V^2)^{0.26}}+1  \right]                
\end{eqnarray}                                         
obtained from a fit to experimental data referred in Ref. \cite{Mara}. The uncertainty in this approximation can be larger than 20--30 $\%$ depending on the $Q^2$ value. It is verified that the effective power $\lambda_e$ is similar for both RC BK (solid line curves) and MPS (long dashed curves) predictions, with the deviation starting only at the higher $Q^2$ values where the predictions differ by a factor 1.5.  This can be a result of the similar small-$x$ behaviour for both models, where the effective power ranges from the soft Pomeron intercept $\lambda_e (Q^2=0)\approx \alpha_{\pom}(0)=1.08$ up to a hard QCD intercept $\lambda_e(Q^2) \simeq cN_c\alpha_s/\pi \approx 0.3$ for large $Q^2$. The data description is fairly good, with the main theoretical uncertainty associated to the choice of the light cone wavefunction (about a 15 $\%$ error).  It was verified that the contribution of real part of amplitude and skewedness are very small for $\rho$ production.


In Fig. \ref{fig:2} we present the  predictions of the RC BK model for the diffractive $J/\Psi$ production and compare with the ZEUS (left panel) and H1  (right panel) data \cite{H1_rho,ZEUS_rho}.  It is verified that the effective power $\lambda_e$ is similar for both RC BK and MPS only in the photoproduction case. The situation changes when the photon virtuality increases. The effective power for RC BK (solid line curves) is enhanced in $Q^2$ in comparison with the non-forward saturation model (long dashed curves).  The data description is reasonable since it is a parameter-free calculation and the uncertainties are similar as for $\rho$ production. For   $J/\Psi$ production, the contribution of real part of amplitude increase by 10 \%  the overall normalization, while the skewedness have a 20 \% effect. In the MPS model, as discussed before, the off-forward effects are absorbed in the parameters of model. The RC BK and MPS predictions  differ by a factor 1.4 for large energies. For sake of completeness, in Fig. \ref{fig:3} the results for $\Upsilon$ photoproduction is presented. The RC BK and MPS predictions are similar in the HERA energy range and differ by a factor 1.5 for large energies. It is known so far that the dipole approach underestimates the experimental data for $\Upsilon$ and the reason is not completely clear \cite{KMW,nos}. However, the deviation concerns only to overall normalization, whereas the energy dependence is fairly described. The referred enhancement in the effective power $\lambda_e$ is already evident in $\Upsilon$ photoproduction as the meson mass, $m_V = 9.46$ GeV, is a scale hard enough for deviations to be present. Skewedness is huge in the $\Upsilon$ case, giving a factor $R_g^2\approx 1.3$ in photoproduction. For this reason, we have included this effect in both models. However, this is not enough to bring the theoretical results closer to experimental measurements.

\begin{figure}[t]
\includegraphics[scale=0.35]{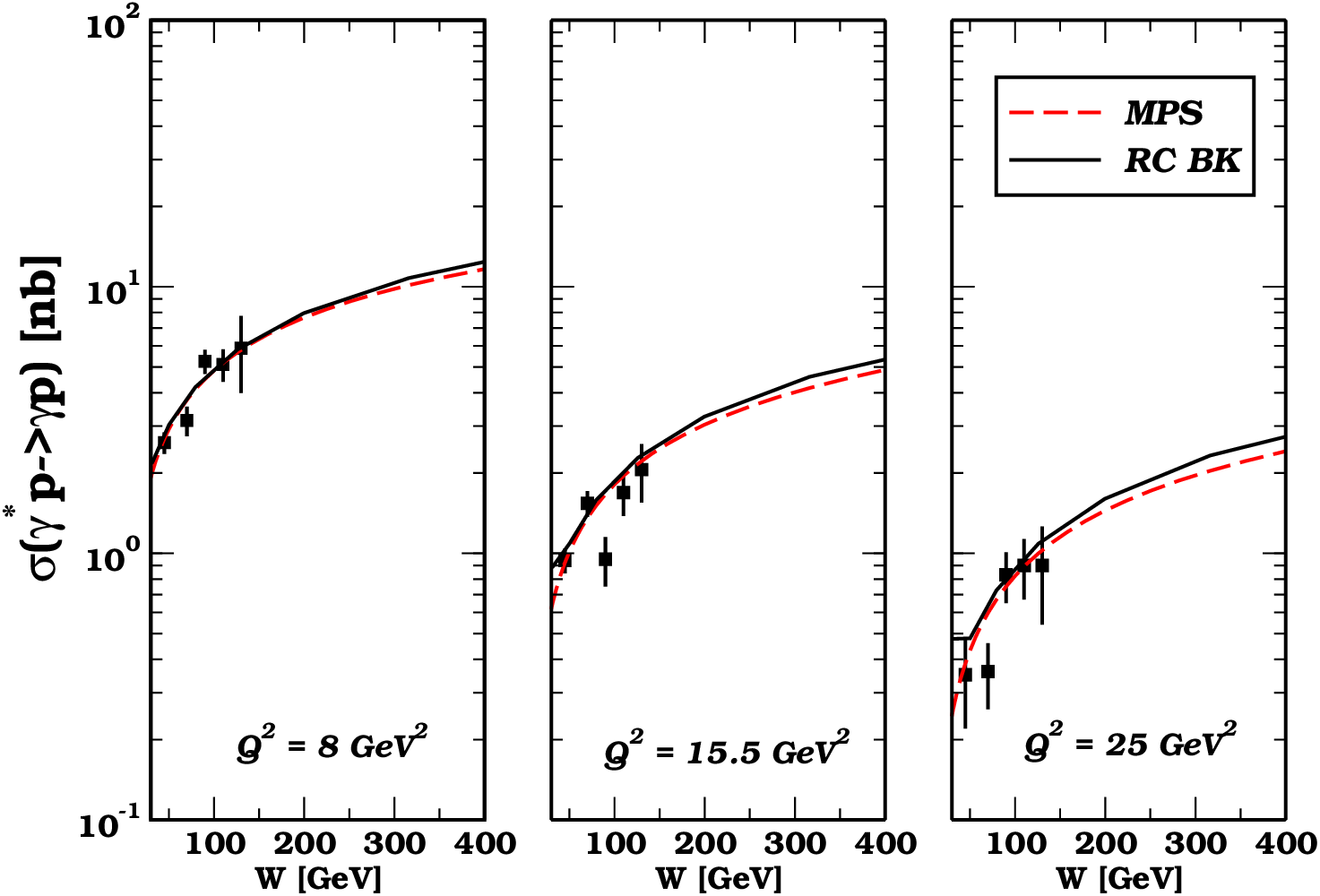}
\caption{(Color online)  Energy dependence of the DVCS cross section for different photon virtualities. Data from   H1 collaboration \cite{H1_dvcs}.}
\label{fig:4}
\end{figure}

Finally, we analyse the DVCS cross section and compare it to the recent H1 data \cite{H1_dvcs}. The cross sections are presented  as a function of $W$, for different values of $Q^2$, in Fig. \ref{fig:4}. Here, the approximations concerning the final state particle are not present and the cross section suffers of less uncertainties. For the slope value, we take the experimental parametrization \cite{H1_dvcs}, $B\,(Q^2)=a[1-b\log(Q^2/Q_0^2)]$, with $a=6.98 \pm 0.54 $ GeV$^2$, $b=0.12 \pm 0.03$ and $Q_0^2 = 2$ GeV$^2$. The situation for DVCS is similar as for vector meson photoproduction, where the effective power $\lambda_e$ is similar for both RC BK and MPS for small virtualities and starts to change as $Q^2$ grows. Skewedness is increasingly important for DVCS at high $Q^2$ and it was introduced for RC BK model. For the MPS model this effect is absorbed in the its parameters as noticed before. The RC BK and MPS predictions are similar for the HERA energy range, describing  the current data, and differ by a factor 1.2 for large energies.

As a summary, we presented a systematic analysis of exclusive production in small-$x$ deep inelastic scattering in terms of the non-linear QCD dynamics. This approach was performed using the recent calculation of the running coupling corrections to the BK equation. In this work we obtained the predictions for the exclusive production of vector mesons and DVCS and compare them to  the available experimental results and the predictions of the MPS model.  The main novelty of this work with respect
to previous phenomenological analyses is the direct use of the running coupling BK
equation to describe the energy and virtuality dependences of exclusive processes at DESY-HERA. We find a
fairly good agreement with experimental data using a parameter-free calculation (parameters are fixed from structure function $F_2$ data). Our main result is that the RC BK evolution equation implies larger cross sections for exclusive processes than  the phenomenological model proposed in \cite{MPS}. Our predictions for both vector meson and DVCS production are relevant for the physics programs in the ongoing experiment LHeC and  in the photoproduction processes in coherent proton - proton interactions at the LHC.


\begin{acknowledgments}
This work was  partially financed by the Brazilian funding
agencies CNPq and CAPES.
\end{acknowledgments}

\end{document}